%% file: main_short.tex
\documentclass[twocolumn,trackchanges]{aastex701}

\newcommand{\momblz}{MoM-BL-z10}

\usepackage{amsmath}
\usepackage{subcaption}

\shortauthors{X.~Cai et al.}

\begin{document}

\title{Discovery of a Broad-Line Active Galactic Nucleus at $z = 9.98$}

\author[0000-0003-3095-7682]{Xingyao Cai}
\affiliation{Department of Pure and Applied Physics, Graduate School of Advanced Science and Engineering, Faculty of Science and Engineering, Waseda University, Shinjuku, Tokyo 169-8555, Japan}
\email[show]{xingyaocai@fuji.waseda.jp}

\author[0000-0002-7779-8677]{Akio K.\ Inoue}
\affiliation{Department of Pure and Applied Physics, Graduate School of Advanced Science and Engineering, Faculty of Science and Engineering, Waseda University, Shinjuku, Tokyo 169-8555, Japan}
\affiliation{Waseda Research Institute for Science and Engineering, Faculty of Science and Engineering, Waseda University, Shinjuku, Tokyo 169-8555, Japan}
\email{akinoue@aoni.waseda.jp}

\author[0000-0003-2984-6803]{Masafusa Onoue}
\affiliation{Waseda Institute for Advanced Study (WIAS), Waseda University, 1-21-1, Nishi-Waseda, Shinjuku, Tokyo 169-0051, Japan}
\affiliation{Kavli Institute for the Physics and Mathematics of the Universe (WPI), The University of Tokyo Institutes for Advanced Study, The University of Tokyo, Kashiwa, Chiba 277-8583, Japan}
\email{masafusa.onoue@aoni.waseda.jp}

\author[0009-0000-6801-1505]{Suzuka Arai}
\affiliation{Department of Pure and Applied Physics, Graduate School of Advanced Science and Engineering, Faculty of Science and Engineering, Waseda University, Shinjuku, Tokyo 169-8555, Japan}
\email{suzukaarai@ruri.waseda.jp}

\author[0000-0001-7440-8832]{Yoshinobu Fudamoto}
\affiliation{Center for Frontier Science, Chiba University, 1-33 Yayoi-cho, Inage-ku, Chiba 263-8522, Japan}
\email{y.fudamoto@chiba-u.jp}

\author[0000-0003-3997-5705]{Rohan P. Naidu}
\affiliation{Institute for Astronomy, University of Hawai`i at M\={a}noa,
2680 Woodlawn Drive, Honolulu, HI 96822, USA}
\email{rnaidu@hawaii.edu}

\author[0000-0001-5851-6649]{Pascal A. Oesch}
\affiliation{Department of Astronomy, University of Geneva, Chemin Pegasi 51, 1290 Versoix, Switzerland}
\affiliation{Cosmic DAWN Center, Niels Bohr Institute, University of Copenhagen, Jagtvej 128, K\o benhavn N, DK-2200, Denmark}
\email{pascal.oesch@unige.ch}

\author[0000-0002-5612-3427]{Jenny E. Greene}
\affiliation{Department of Astrophysical Sciences, Princeton University, 4 Ivy Lane, Princeton, NJ 08544, USA}
\email{jgreene@astro.princeton.edu}

\author[0000-0002-2057-5376]{Ivo Labbe}
\affiliation{Centre for Astrophysics and Supercomputing, Swinburne University of Technology, Melbourne, VIC 3122, Australia}
\email{ilabbe@swin.edu.au}

\author[0000-0003-2871-127X]{Jorryt Matthee}
\affiliation{Institute of Science and Technology Austria (ISTA), Am Campus 1, 3400 Klosterneuburg, Austria}
\email{jorryt.matthee@ista.ac.at}

\author[0000-0002-2380-9801]{Anna de Graaff}
\affiliation{Max-Planck-Institut f\"ur Astronomie, K\"onigstuhl 17, D-69117 Heidelberg, Germany}
\email{degraaff@mpia.de}

\author[0000-0001-6958-7856]{Yuma Sugahara}
\affiliation{Department of Natural Sciences, Faculty of Science and Engineering, Tokyo City University, 1-28-1 Tamazutsumi, Setagaya, Tokyo 158–8557, Japan}
\email{sugayu@aoni.waseda.jp}

\author[0000-0003-4985-0201]{Ken Mawatari}
\affiliation{Faculty of Industrial Technology, Tsukuba University of Technology, 4-3-15 Amakubo, Tsukuba, Ibaraki 305-8520, Japan}
\email{mawatari@aoni.waseda.jp}

\author[0000-0001-8460-1564]{Pratika Dayal}
\affiliation{Canadian Institute for Theoretical Astrophysics, 60 St George St, University of Toronto, Toronto, ON M5S 3H8, Canada}
\affiliation{David A. Dunlap Department of Astronomy and Astrophysics, University of Toronto, 50 St George St, Toronto ON M5S 3H4, Canada }
\affiliation{Department of Physics, 60 St George St, University of Toronto, Toronto, ON M5S 3H8, Canada}
\email{pdayal@cita.utoronto.ca}

\author[0000-0002-5458-4254]{Xi Kang}
\affiliation{Institute for Astronomy, School of Physics, Zhejiang University, Hangzhou 310058, Zhejiang, China}
\affiliation{Center for Cosmology and Computational Astrophysics, Zhejiang University, Hangzhou 310058, Zhejiang, China}
\email{kangxi@zju.edu.cn}

\begin{abstract}
The James Webb Space Telescope (JWST) has uncovered a previously unexplored population of faint active galactic nuclei (AGNs) in the early Universe.
We present \momblz{} at $z=9.9837\pm0.0009$, identified through a systematic search for broad-line AGNs at $z\geq6$ using public JWST/NIRSpec PRISM spectra from the DAWN JWST Archive.
The PRISM spectrum exhibits a broad component in H$\beta$ at $5.35~\micron$.
A narrow$+$broad Gaussian fit yields a broad H$\beta$ detection at ${\rm S/N}=3.87$, with ${\rm FWHM}=3480^{+2040}_{-950}$ km s$^{-1}$ and $L_{{\rm H}\beta,{\rm broad}}=1.25^{+0.33}_{-0.31}\times10^{42}$ erg s$^{-1}$.
The detection of [Ne~{\sc iii}] $\lambda3870$ and the high [Ne~{\sc iii}]/[O~{\sc ii}] and [O~{\sc iii}]/H$\beta$ ratios further support the AGN interpretation.
Its rest-frame UV and optical continuum slopes, compact NIRCam morphology, and a tentative MIRI F770W detection suggest similarities to Little Red Dots (LRDs).
The broad H$\beta$ detection enables the first virial SMBH mass estimate at $z\sim10$.
We derive $M_{\rm BH}=3.6^{+5.8}_{-1.7}\times10^7~M_\odot$ and $\lambda_{\rm Edd}=0.11^{+0.10}_{-0.07}$ using the standard single-epoch virial calibration and quasar bolometric correction.
We also consider an alternative scenario adopting the LRD-specific bolometric correction and assuming $\lambda_{\rm Edd}=1$, which yields $M_{\rm BH}=2.1^{+0.6}_{-0.5}\times10^6\ M_\odot$.
Under either scenario, the inferred SMBH mass at a cosmic age of only 470 Myr highlights the challenge of rapid BH growth in the early Universe, suggesting super-Eddington accretion from a stellar-mass seed or a more massive seed.
As the highest-redshift broad-line AGN identified to date, \momblz{} provides a new observational benchmark for understanding the early growth of supermassive black holes.
\end{abstract}


\section{Introduction}\label{sec:introduction}

The existence of luminous quasars powered by billion-solar-mass black holes within the first billion years poses a major challenge to our understanding of the early growth history of SMBHs \citep{fan2023quasars}.
Wide-field optical and near-infrared surveys have identified hundreds of luminous quasars down to $z=7.8$ \citep[e.g.,][]{banados2018j1342,yang2020j1007,wang2021j0313, yang2026}.
These luminous quasars place stringent constraints on seed BH formation and the rapid accretion histories of the rarest and most extreme population.
However, those identified in wide-field surveys are limited to the bolometric luminosity range of $L_{\rm bol}\gtrsim 10^{46}$ erg~s$^{-1}$ \citep{onoue2019shellq,willott2010}.
Fainter and therefore less massive populations of active galactic nuclei (AGNs) are essential to investigate the buildup of the more representative SMBH population in the early universe.

The unprecedented infrared sensitivity of the James Webb Space Telescope \citep[JWST;][]{rigby2023} enables us to probe the luminosity regime of Seyfert galaxies ($L_{\rm bol} \gtrsim10^{44}~{\rm erg~s^{-1}}$) at high redshift.
This new high-redshift AGN population has been identified via spectroscopic detection of broad Balmer emission lines \citep[e.g., ][]{onoue2023,kocevski2023hidden,harikane2023firstcensus,kokorev2023lrd,maiolino2024jades,juodzbalis2026jades}.
Single-epoch BH mass estimates of these populations yield $M_{\rm BH}\sim10^6$--$10^8~M_\odot$, which are more than an order of magnitude lower than those of pre-JWST studies.

Remarkably, a non-negligible fraction of the JWST-selected AGNs exhibits compact morphology with ``V''-shape continuum colors between rest-frame UV and optical wavelengths.
These objects, called Little Red Dots \citep[LRDs; e.g.,][]{matthee2024lrd,greene2024lrd}, often exhibit broad Balmer lines, indicating the presence of accreting BHs.
However, alternative stellar and wind-based interpretations have also been proposed \citep{chisholm2026lrd,nandal2026lrd,zwick2026lrd,martins2026lrd,naidu2026lrdengine}.
Proposed explanations for their characteristic spectra include dust-obscured AGNs \citep{noboriguchi2023,greene2024lrd}, compact galaxies dominated by dusty star formation \citep{akins2023lrd,akins2025lrd} or evolved stellar populations \citep{baggen2024lrd}, and accreting BHs embedded in dense gas envelopes \citep[``BH star'' model; ][]{inayoshi2025densemass, taylor2025capers, naidu2025bh*,torralba2026bh*,degraaff2025lrd}.

Extensive efforts have been made to explore this LRD population out to $z\sim10$.
Broad H$\beta$ emission has been detected in several LRDs, including UNCOVER~20466 at $z=8.50$ \citep{kokorev2023lrd}, CANUCS-LRD-z8.6 at $z=8.63$ \citep{tripodi2025canucs}, and CAPERS-LRD-z9 at $z=9.288$ \citep{taylor2025capers}, indicating that these sources host broad-line (BL) AGNs.
\citet{tanaka2025} reported a photometric candidate for a $z\sim10$ LRD.
There are also a few AGN candidates at similar redshifts.
UHZ1 at $z=10.1$ was initially reported to show a hard X-ray excess \citep{bogdan2023bhmassevolution} and was subsequently spectroscopically confirmed at $z_{\rm spec}=10.073$ \citep{goulding2023uhz1,zou2026uhz1}.
GN-z11 at $z=10.6$ was first identified as a Lyman-break galaxy \citep{Oesch16}, and its possible AGN nature has been discussed based on rest-frame UV/optical emission-line diagnostics \citep{maiolino2024gnz11}.
However, neither UHZ1 nor GN-z11 has a securely detected broad emission-line component in its JWST spectrum, and the nature of the putative AGN activity in these sources remains debated \citep{alvarez2025gnz11, zou2026uhz1,chen2026spurs}.

In this Letter, we report the discovery of \momblz{}, a BL AGN with spectroscopic redshift of $z=9.9837$.
This object is identified by a systematic archival search of publicly available JWST/NIRSpec PRISM data for BL AGNs independently of imaging colors or compactness. We subsequently assess its LRD classification using the spectrum, NIRCam morphology and MIRI photometry.
We adopt a flat $\Lambda$CDM cosmology with $H_0=70~{\rm km~s^{-1}~Mpc^{-1}}$, $\Omega_{\rm M}=0.3$, and $\Omega_\Lambda=0.7$, and use AB magnitudes \citep{oke1983ABmag}.
All emission-line wavelengths are quoted in vacuum.

\section{Data and Analysis}
\label{sec:dataandanalysis}

Here we briefly describe our BL AGN search.
A full description of the complete sample will be presented in X.~Cai et al.\ in preparation.

We use the public Near-Infrared Spectrograph (NIRSpec; \citealt{jakobsen2022jwstdoc}) PRISM spectra from the DAWN JWST Archive \citep[DJA version 4.4;][]{valentino2023atlasDJA, heintz2024strongDJA,degraaff2025rubiesDJA}\footnote{\href{https://dawn-cph.github.io/dja/}{https://dawn-cph.github.io/dja/}}.
After removing the duplicated observations, the parent sample comprises 1,247 sources with grade-3 robust spectroscopic redshifts at $6.0\leq z_{\rm spec}\leq 10.3$, ensuring that H$\beta$ falls within NIRSpec spectral coverage.
We do not apply any criteria based on photometric colors and compactness for the parent selection.
To account for possible changes in continuum slopes associated with LRD-like spectra, we adopt a broken power-law model with a fixed break at $\lambda_{\rm br}=3645$~\AA.
We use 19 fitting windows spanning rest-frame 1275--8150~\AA, following \texttt{PyQSOFit} \citep{guo2018pyqsofit,shen2019windows} with slight modifications (X.~Cai et al.\ in preparation).
After subtracting the continuum, we compare narrow-only and narrow$+$broad Gaussian models of H$\beta$+[O~{\sc iii}]$\lambda\lambda4960,5008$ complex and, when covered, H$\alpha$+[N~{\sc ii}]$\lambda\lambda6550,6585$ complex. The emission-line models are convolved with the wavelength-dependent NIRSpec line spread function (LSF) of \citet{jakobsen2022jwstdoc}, and adopt an effective resolving power 1.4 times the pre-launch estimate \citep{degraaff2024lsf}.

We use the Bayesian Information Criterion (BIC) \citep{schwarz1978bicdefinition} to select BL AGN candidates.
The selection threshold is $\Delta\mathrm{BIC} \equiv \mathrm{BIC}_{\rm narrow+broad} - \mathrm{BIC}_{\rm narrow-only} < -6$,
which is commonly interpreted as strong evidence for the narrow$+$broad model over the narrow-only model \citep{kass1995bayesfactors}.
This threshold is also consistent with recent JWST BL AGN searches \citep{greene2024lrd,matthee2024lrd,kocevski2023hidden,maiolino2024jades}.
We refine the model fit to the emission-line complex for BIC-selected candidates using \texttt{emcee} Markov chain Monte Carlo sampling \citep{foremanmackey2013emcee}, adding a 5\% absolute flux-calibration uncertainty in quadrature to the statistical errors \citep{gordon2022jwstflux,boker2023jwstflux} and adopt the posterior medians and 16th--84th percentiles.
We define the BL S/N as the median flux divided by the mean of its lower and upper 68\% uncertainties; requiring S/N $\geq3$ leaves 66 candidates. \momblz{} satisfies these criteria and is investigated below.
Full details of the selection and the BL AGN catalog will be presented in a companion paper (X.~Cai et al.\ in preparation).

\section{\momblz{}: a $\MakeLowercase{z}=10$ broad-line AGN}
\label{sec:momblz}

\input{mom_blz10_properties}

\begin{figure*}
\centering
\includegraphics[width=\textwidth, trim={25 10 0 0}, clip]{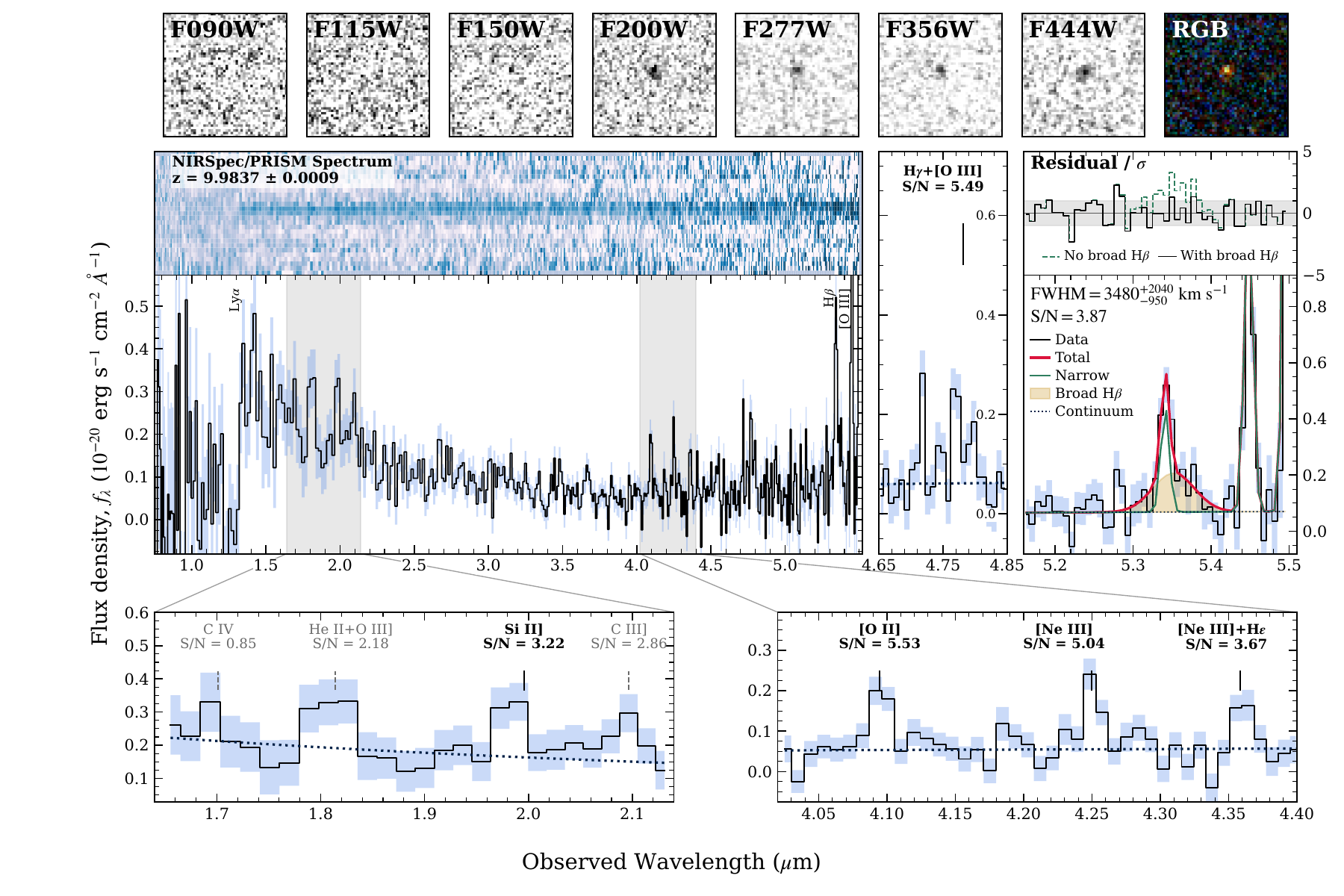}
\caption{JWST spectroscopy and imaging  of \momblz.
The top row shows $2\times2$ arcsec$^2$ cutout images in 7 NIRCam broadbands from F090W to F444W.
The rightmost panel shows a pseudo-color image (blue: F090W+F115W+F150W; green: F200W+F277W; red: F356W+F444W).
The middle panels show the 2D and 1D NIRSpec/PRISM spectra.
In the 2D spectrum, positive and negative signals are shown in blue and white, respectively.
The blue shaded region in the 1D spectrum indicates $1\sigma$ uncertainties.
The positions of Ly$\alpha$ and H$\beta$+[\ion{O}{3}] are marked.
The right-hand panels show the H$\gamma$+[O~{\sc iii}] $\lambda4364$ and the H$\beta$+[O~{\sc iii}] complex regions.
In the H$\beta$+[O~{\sc iii}] panel, we show the narrow+broad Gaussian fit to H$\beta$.
The upper inset shows the normalized residuals for fits with and without the broad H$\beta$ component.
The bottom panels show zoomed-in views of weaker emission-line features, with labels in black for detections with S/N $\geq3$ and in gray for features with S/N $<3$.
}
\label{fig:spectra}
\end{figure*}

\subsection{Emission Line Measurements}
\label{sec:momblz_spec}

\momblz{} is the highest-redshift source in our catalog.
The NIRSpec MSA PRISM/CLEAR observations were obtained as part of the ``Mirage or Miracle'' (MoM) program (GO~5224; PIs: P.A.~Oesch and R.P.~Naidu; \citealt{naidu2026mom}), with a total effective exposure time of 15,756~s.
Figure~\ref{fig:spectra} shows its NIRSpec and NIRCam data (Section~\ref{sec:momblz_phot}).
\momblz{} exhibits a broad H$\beta$ emission line at $5.35$~$\mu$m, which is beyond the cut-off wavelength of the standard JWST pipeline (5.3~$\mu$m).
This detection was made possible by the DJA data reduction that recovers second and third-order spectra \citep{valentino2025, pollock2026auroralline}.
We obtain the systemic redshift of $z_{\rm spec}=9.9837\pm0.0009$ by jointly fitting narrow optical emission lines including [O~{\sc ii}] $\lambda\lambda3727,3730$, [Ne~{\sc iii}] $\lambda\lambda3870,3969$, and [O~{\sc iii}] $\lambda4960$.
We note that only the blue edge of [O~{\sc iii}] $\lambda5008$ is covered by the PRISM data, and we do not use it in our analysis.

The presence of a broad component in H$\beta$ is statistically preferred, as the narrow$+$broad line model returns a smaller BIC relative to the narrow-line-only model ($\Delta{\rm BIC}=-15.55$).
The decomposed narrow and broad H$\beta$ fluxes are $F_{\rm H\beta,broad}=93.4^{+24.9}_{-23.3}\times10^{-20}\,{\rm erg\,s^{-1}\,cm^{-2}}$ and $F_{\rm H\beta,narrow}=51.9^{+10.0}_{-10.3}\times10^{-20}\,{\rm erg\,s^{-1}\,cm^{-2}}$, respectively.
The corresponding detection significance of the broad line is S/N $=3.87$.
The intrinsic broad H$\beta$ width is ${\rm FWHM}=3480^{+2040}_{-950}~{\rm km\,s^{-1}}$, after correcting for instrumental broadening.
There is a marginal velocity redshift of $604^{+449}_{-429}\,{\rm km\,s^{-1}}$ relative to the narrow lines. Here positive and negative velocity offsets indicate redshifted and blueshifted, respectively.
The best-fit emission-line model and the reduction in the fit residuals achieved by including the broad H$\beta$ component are shown in the rightmost middle panel of Figure~\ref{fig:spectra}.

For weaker emission lines, we measure their fluxes by integrating the continuum-subtracted spectrum over a wavelength window twice the width of the instrumental LSF, centered at the expected wavelengths based on the systemic redshift.
Adopting a detection threshold of S/N $\geq 3$, we detect the H$\gamma$+[O~{\sc iii}] $\lambda4364$ complex (S/N $=5.49$), Si~{\sc ii}] $\lambda1817$ (3.22), [O~{\sc ii}] $\lambda3728$ (5.53), [Ne~{\sc iii}] $\lambda3869$ (5.04), and [Ne~{\sc iii}] $\lambda3968$+H$\epsilon$ (3.67).
Their measured fluxes are listed in Table~\ref{tab:mom_bl_z10_properties}.
We note that, for H$\gamma$+[O~{\sc iii}], we report the total flux over $4.755$--$4.807$~\micron.
The narrow spike at $\approx4.72~\micron$ is not included in this wavelength window.
We do not report the line profiles of these weak lines, given their low detection significance and the low spectral resolution of the PRISM data.
We also find tentative evidence for C~{\sc iii}] $\lambda1909$ (S/N $=2.86$) and the He~{\sc ii} $\lambda1640$+O~{\sc iii}] $\lambda1663$ complex (2.18), despite their low detection significance.
Other major AGN emission lines such as C~{\sc iv} $\lambda1549$ and Mg~{\sc ii} $\lambda2798$ are not significantly detected.

We highlight that \momblz{} at $z=9.98$ represents the highest-redshift broad-line AGN currently known, beyond CAPERS-LRD-z9 at $z=9.29$ \citep{taylor2025capers}, while luminous quasars are known down to $z=7.77$ \citep{yang2026}.
The redshift of \momblz{} approaches the upper limit for detecting broad H$\beta$ with the extended NIRSpec wavelength coverage \citep{valentino2025}.
Broad-line detection in higher-order Balmer emission lines would be more challenging (H$\gamma$/H$\beta$=0.53; \citealt{juodzbalis2026jades}).
At $z\gtrsim10$, AGN identification has largely relied on high-ionization or high-density emission-line diagnostics \citep[e.g., GN-z11;][]{maiolino2024gnz11, alvarez2025gnz11} or  X-ray detections \citep[e.g., UHZ1;][]{goulding2023uhz1, bogdan2023bhmassevolution}.
However, GN-z11 at $z=10.6$ does not show a dominant broad-H$\alpha$ component in its MIRI/MRS spectrum \citep{alvarez2025gnz11}. Its optical emission is consistent with a metal-poor starburst, although a weak AGN contribution cannot be excluded \citep{alvarez2025gnz11}.
For UHZ1, recent work also favors a low-mass, metal-poor, star-forming galaxy interpretation rather than a luminous obscured AGN \citep[see e.g.][]{zou2026uhz1}.
There is a photometric candidate for a $z\sim10$ LRD, CW-LRD-z10, reported in \citet{tanaka2025}, with MIRI follow-up spectroscopy scheduled under JWST-GO-9483.

\begin{figure*}[t]
    \centering
    \includegraphics[width=0.8\textwidth, trim={10 10 10 0}, clip]{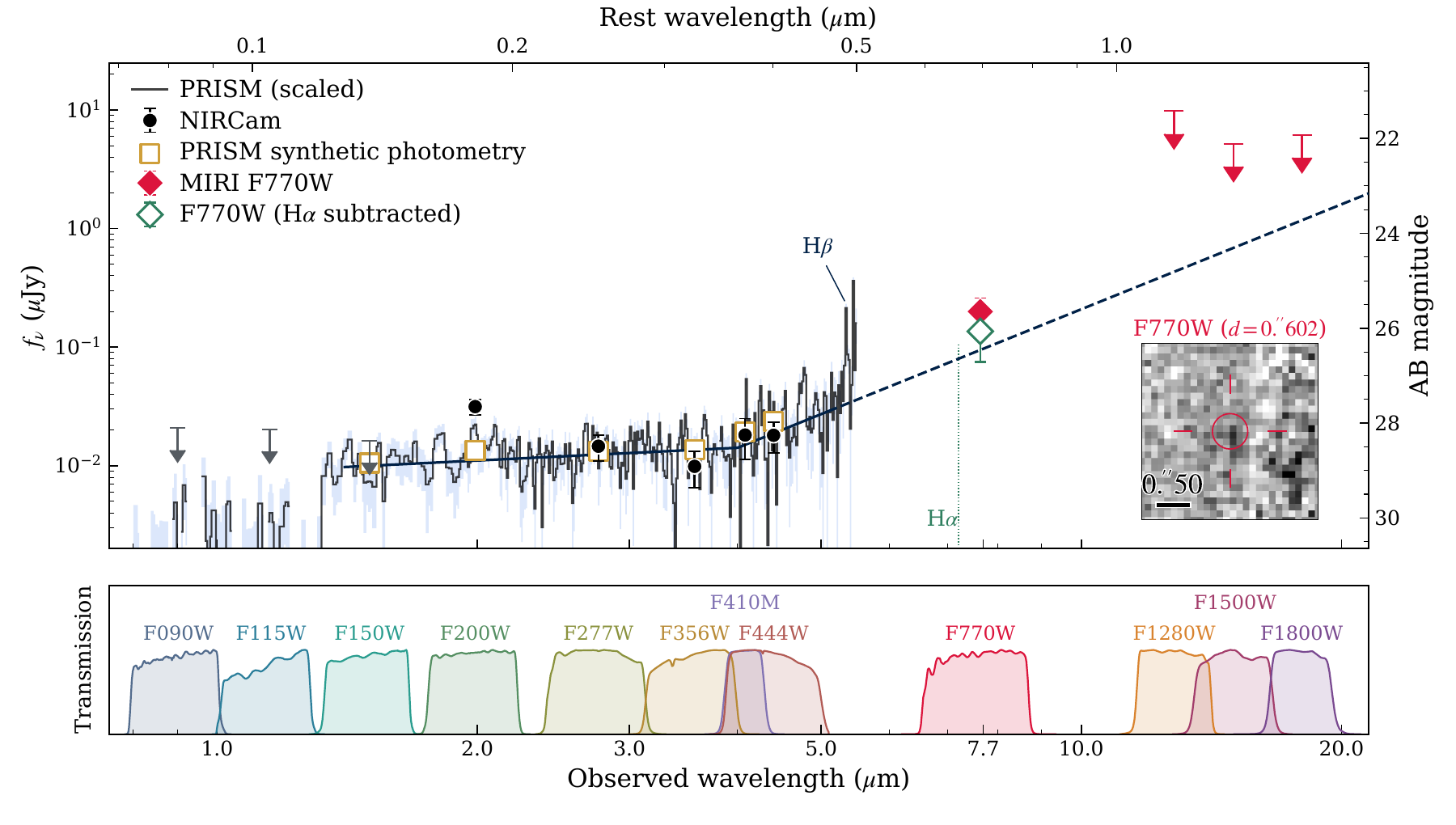}
    \caption{Rest-frame UV-to-near-infrared SED of \momblz{}, combining NIRCam and MIRI photometry with a scaled PRISM spectrum (see text).
    For visualization, the PRISM spectrum is resampled onto a logarithmic wavelength grid with a fixed velocity step of $\Delta v=1160~{\rm km~s^{-1}}$.
    Black circles show the NIRCam photometry, and grey arrows indicate the  $3\sigma$ upper limits for undetected bands.
    Open orange squares show synthetic PRISM photometry, obtained by convolving the scaled PRISM spectrum with the NIRCam filter transmission curves.
    The filled diamond shows the tentative MIRI F770W detection.
    The open green diamond shows the estimated continuum flux after subtracting H$\alpha$ contribution, assuming $L_{{\rm H}\alpha}=10L_{{\rm H}\beta}$.
    Red arrows indicate the $3\sigma$ upper limits of other MIRI bands.
    The thick solid  lines show the broken-power-law continuum model, while the dashed line is the extrapolation to the MIRI coverage.
    The inset  shows a $3\times3$ arcsec$^2$ F770W cutout image, where a red circle indicates the photometric aperture ($d=0\farcs602$) centered on the source position.
    The transmission curves of the NIRCam and MIRI filters  are shown in the bottom panel.
    }
    \label{fig:miri}
\end{figure*}

\subsection{Continuum Measurements}
\label{sec:momblz_phot}

Multiple JWST programs (PRIMER, GO~1837, PI: J.~Dunlop \citep{donnan2024primer}; BEACON, GO~3990, PI: T.~Morishita \citep{morishita2025beacon} and GO~5324, PI: J.~Pierel) observed \momblz{} with NIRCam.
Seven broadband images are available in the DJA archive from F090W through F444W, which we present in the top panels of Figure~\ref{fig:spectra}.
The source has a compact morphology, with an F444W aperture flux ratio of $f_{\rm F444W}(0\farcs50)/f_{\rm F444W}(0\farcs25)=1.23\pm0.15$, measured using aperture diameters of $0\farcs50$ and $0\farcs25$.
This satisfies the compactness criterion of $f_{\rm F444W}(0\farcs50)/f_{\rm F444W}(0\farcs25)<1.7$ adopted in \citet{rinaldi2025companion}.
We also perform independent aperture photometry using a $0\farcs20$-diameter aperture to find $f_{\rm F444W}(0\farcs50)/f_{\rm F444W}(0\farcs20)=1.61\pm0.21$.
This satisfies the compactness criterion of $1.4$--$2.0$ adopted by \citet{tanaka2025} for their $z>10$ LRD selection.

Our continuum model based on the PRISM spectrum yields the rest-frame $1450~{\rm \AA}$ absolute magnitude of  $M_\mathrm{ UV}=-19.37^{+0.14}_{-0.13}$, which is similar to those of $z>9$ BL AGNs (e.g., $M_{\rm UV}=-19.3$ for CANUCS-LRD-z9; \citealt{Morishita2026AGN}), and about six magnitudes fainter than those of the luminous $z\sim7.5$ quasars \citep[$M_\mathrm{1450}\sim-26$; ][]{banados2018j1342, yang2020j1007, wang2021j0313}.

\momblz{} exhibits a characteristic continuum shape in the rest-frame ultraviolet and optical wavelengths with the PRISM-based continuum slope measurements yielding $\beta_{\rm UV}=-1.64\pm0.21$ and $\beta_{\rm opt}=0.94\pm0.83$.
Although the optical slope is poorly constrained by the limited rest-frame wavelength coverage of the $z\approx10$ source, at face value it satisfies the criteria for LRDs adopted in \citet{degraaff2025lrd}.

We further investigate the rest-frame optical-and-infrared continuum colors with existing Mid-Infrared Instrument \citep[MIRI;][]{rieke2015miri} imaging datasets in the target field, obtained through the PRIMER \citep[GO~1837, PI: J.~Dunlop;][]{donnan2024primer,perezgonzalez2024primer}  and MINERVA \citep[GO~7814, PI: A.~Muzzin;][]{muzzin2025minerva} programs.
The PRIMER observations cover the F770W and F1800W filters, with exposure times of 1,776~s and 1,787~s, respectively.
The MINERVA observations provide F1280W and F1500W imaging, each with an exposure time of 760~s.

We report the multi-band photometry in Table~\ref{tab:mom_bl_z10_properties}.
Figure~\ref{fig:miri} presents the NIRCam and MIRI photometric SED of \momblz, together with the NIRSpec PRISM spectrum.
For NIRCam, we adopt the photometry measured within a $0\farcs5$-diameter aperture from the DJA catalog \citep{valentino2023atlasDJA}.
Three-sigma upper limit fluxes are provided for the bands with $\mathrm{S/N} \lesssim 3$.
For MIRI, we perform aperture photometry using aperture diameters of $0\farcs602$ (F770W), $0\farcs860$ (F1280W), $1\farcs012$ (F1500W), and $1\farcs190$ (F1800W).
These aperture diameters are set to twice the PSF FWHM in each band\footnote{\url{https://jwst-docs.stsci.edu/jwst-mid-infrared-instrument/miri-performance/miri-point-spread-functions}}.
We estimate the background noise in the MIRI images using random apertures of the same size as those adopted for the source photometry, placed within a $10\times10$ arcsec$^2$ region centered on the source.
As \momblz{} is located close to the boundaries of the MIRI imaging coverage, we mask pixels affected by edge artifacts in the background estimate.
In the F1280W and F1500W images, the source is covered by fewer dither positions than the central regions, resulting in a shorter effective exposure time.
To account for this reduced imaging depth, we restrict the random apertures to regions with the same dither coverage as the source.
As a result, we find a tentative detection of \momblz{} in F770W with $f_\nu({\rm F770W})=0.20\pm0.06~\mu$Jy (S/N $=3.33$), while it is not significantly detected in the three redder filters.
The $3\sigma$ upper limits in F1280W, F1500W, and F1800W are 9.79, 5.20 and 6.18~$\mu$Jy, respectively.

We now revisit the continuum properties of \momblz{} using the additional constraints from MIRI photometry, although the F770W detection remains tentative.
The measured NIRCam+MIRI colors, ${\rm F277W}-{\rm F444W}=0.23\pm0.42$ and ${\rm F444W}-{\rm F770W}=2.61\pm0.45$~mag, are consistent with the expected colors of $z\sim10$ LRDs.
\citet{tanaka2025} adopt color cuts of $0<{\rm F277W}-{\rm F444W}<3.0$ and ${\rm F444W}-{\rm F770W}>1.5$ to select their $z>10$ LRD candidates, both of which are satisfied by \momblz{}.
Together with its relatively blue UV continuum slope and compact morphology (Section~\ref{sec:momblz_phot}), these colors suggest that \momblz{} exhibits characteristics consistent with being a $z\approx10$ LRD.
Nevertheless, we refrain from classifying \momblz{} as an LRD from the current dataset, given the limited constraints on the rest-frame optical continuum shape and its low-detection significance in F770W (S/N $=3.33$).
Deeper NIRSpec data and/or MIRI imaging are needed to better constrain its continuum shape and establish its nature.

\subsection{H$\alpha$ Contribution to F770W}
\label{sec:momblz_Ha}

Redshifted H$\alpha$ emission of \momblz{} falls at $7.21~\mu$m, thus contributing to MIRI F770W photometry.
To quantify this contribution, we first normalize the PRISM spectrum to the NIRCam photometry by comparing the synthetic broadband fluxes derived from the PRISM spectrum with the observed NIRCam fluxes.
This yields a scaling factor of $s=0.509\pm0.063$ (i.e., the PRISM spectrum is $\approx2$ times higher).
Applying this scaling factor to the observed H$\beta$ flux and assuming a Balmer decrement of $L_{{\rm H}\alpha}/L_{{\rm H}\beta}=10$, as typically observed in LRDs \citep[e.g., $H\alpha/H\beta=10.4$ in][]{lin2026bridgelrd,torralba2026scatterwings,degraaff2025lrd,kiyota2026lrddecrement}, we estimate an H$\alpha$ contribution of $0.064\pm0.012~\mu$Jy to F770W.
Subtracting this contribution from the observed F770W photometry yields a residual flux density of $0.136\pm0.061~\mu$Jy.
The H$\alpha$-subtracted F770W flux, shown as an open green diamond in Figure~\ref{fig:miri}, is consistent with the extrapolation of the single power-law model fitted to the rest-frame optical continuum of the PRISM spectrum.
The marginal F770W detection is therefore consistent with the presence of a red rest-frame optical continuum in \momblz.

\begin{figure*}[t!]
    \centering
    \includegraphics[width=0.8\linewidth, trim={0 10 0 0}, clip]{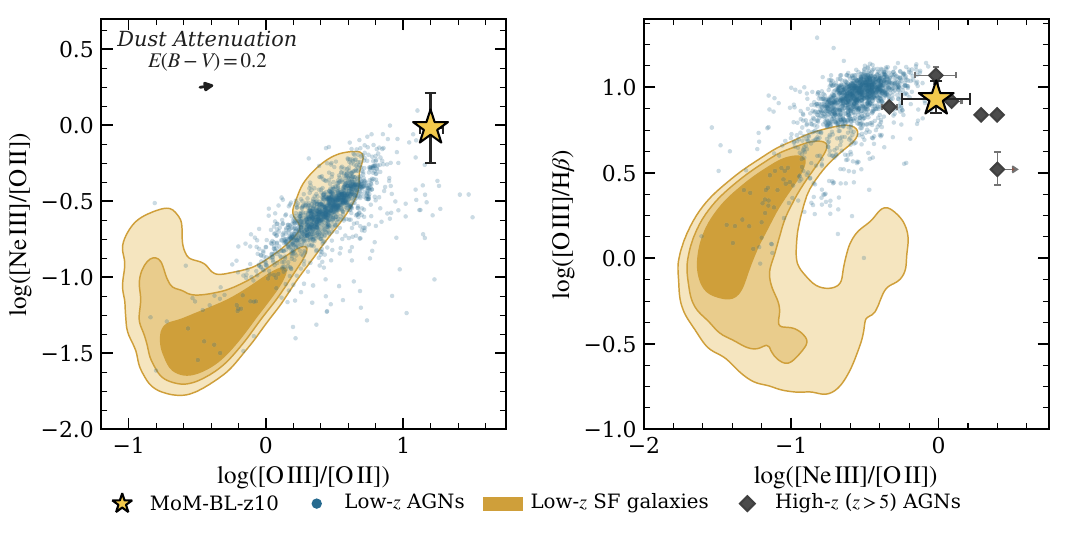}
    \caption{The [O~{\sc ii}]-FLR (left) and OHNO (right) diagnostics. 
    The yellow star shows \momblz{}, where we infer the [O~{\sc iii}] $\lambda5008$ from the [O~{\sc iii}] $\lambda4960$ flux by assuming a fiducial flux ratio of 2.98,  and only the narrow H$\beta$ component is used.
    The error bars reflect the propagated $1\sigma$ measurement uncertainties of the emission-line fluxes.
    An additional 0.20-dex uncertainty is added in quadrature to $\log([\mathrm{Ne~III}]/[\mathrm{O~II}])$ to account for possible blending with H$\eta$, H$\zeta$, and \ion{He}{1}~$\lambda3889$ \citep[e.g.,][]{tripodi2025canucs}.
    Blue points show the SDSS AGN sample selected from BPT and [Ne\,{\sc v}] $\lambda3426$ emission in the Portsmouth SDSS galaxy-property catalogue \citep{feuillet2024lines}; orange contours indicate 50, 70, and 90\% of the density of low-redshift BPT-selected MPA--JHU SF galaxies from the MPA--JHU catalogue of SDSS DR8 galaxy spectra \citep{brinchmann2004sf,kauffmann2003bpt}.
    Other symbols represent high-redshift ($z>5$) JWST AGNs published in the literature \citep{trussler2023ohno,tripodi2024ohno,ubler2024ohno,nandal2025ohno,tripodi2025canucs,backhaus2024ceersohno}.
    The arrow in the left panel indicates the effect of dust attenuation corresponding to $E(B-V)=0.2$ mag, assuming the attenuation law of \citet{calzetti2000dustcurve}.
    }
    \label{fig:ohno}
\end{figure*}

\section{Discussion} \label{sec:discussion}
\subsection{Emission Line Diagnostics} \label{sec:emission_lines}

The metal-line excitation provides a complementary check on the AGN interpretation of \momblz.
We obtain $\log([\mathrm{Ne\,III}]/[\mathrm{O\,II}])=-0.016^{+0.231}_{-0.233}$, $\log([\mathrm{O\,III}]/[\mathrm{O\,II}])=1.199^{+0.091}_{-0.079}$, and $\log([\mathrm{O\,III}]/\mathrm{H}\beta_{\rm narrow})=0.933^{+0.103}_{-0.084}$.
Here [Ne~{\sc iii}] denotes $\lambda3870$, [O~{\sc ii}] denotes the unresolved $\lambda\lambda3727,3730$ doublet, and [O~{\sc iii}]$\lambda5008$ is inferred from [O~{\sc iii}]$\lambda4960$ using the theoretical ratio of 2.98 \citep{storey2000OIIIratio}. We propagate measurement errors and add an additional 0.20-dex [Ne~{\sc iii}] blending uncertainty to the [Ne~{\sc iii}]/[O~{\sc ii}] ratio. This term accounts for possible contamination of [Ne~{\sc iii}] by H$\eta$, H$\zeta$, and He~{\sc i}~$\lambda3889$ at PRISM resolution, following the discussion of \citet{tripodi2025canucs}.

Figure~\ref{fig:ohno} compares \momblz{} in the [Ne~{\sc iii}]/[O~{\sc ii}] versus [O~{\sc iii}]/[O~{\sc ii}] ([O~{\sc ii}]-forbidden line ratio, FLR) and [O~{\sc iii}]/H$\beta$ versus [Ne~{\sc iii}]/[O~{\sc ii}] (OHNO) diagrams.
The AGN reference comprises SDSS galaxies from \citet{feuillet2024lines}, selected by BPT AGN classification and [Ne~{\sc v}] S/N $>3$. The SF galaxies reference comprises around 40,000 MPA--JHU DR8 star-forming spectra \citep{brinchmann2004sf, thomas2013sdssgalaxies}\footnote{\href{https://www.sdss4.org/dr12/spectro/galaxy_portsmouth/}{https://www.sdss4.org/dr12/spectro/galaxy\_portsmouth/}} with BPT lines below the \citet{kauffmann2003bpt} [N~{\sc ii}]/H$\alpha$ boundary.

\momblz{} lies in the AGN-like region in both diagnostics. In the [O~{\sc ii}]-FLR diagram (left panel of Figure~\ref{fig:ohno}), high ratios in both axes can reflect a high ionization parameter or suppression of [O~{\sc ii}] in dense gas \citep{feuillet2024lines}. \momblz{} is beyond the SF contours, towards the high-excitation end of the low-$z$ SDSS AGN distribution. The OHNO diagram (right panel of Figure~\ref{fig:ohno}) instead compares [O~{\sc iii}] with narrow H$\beta$. \momblz{} also lies in the AGN distribution. For $E(B-V)=0.2$, the \citet{calzetti2000dustcurve} dust correction still leaves \momblz{} outside of SF contours, so we exclude the dust attenuation as the explanation for its separation from the SF locus. We therefore identify \momblz{} as an AGN based on the combined evidence from its broad H$\beta$ and its position in both line-ratio diagnostics.

\subsection{Black Hole Mass Estimate} \label{sec:BHmass}

The broad H$\beta$ detection of \momblz{} enables us to estimate its central black hole mass.
We first estimate its virial black hole mass using the single-epoch calibration of \citet{kaspi2000bhmass}, based on rest-frame 5100~\AA\ continuum luminosity ($L_{5100}$) and the broad H$\beta$ FWHM.
The observed continuum luminosity from the PRISM spectrum, without correction for possible host galaxy contamination and inconsistency between the NIRCam and NIRSpec data (Sec.~\ref{sec:momblz_Ha}), is $L_{5100}=5.4^{+1.5}_{-1.2}\times10^{43}~{\rm erg~s^{-1}}$.
Combining this with the measured broad H$\beta$ ${\rm FWHM}=3480^{+2040}_{-950}~{\rm km\,s^{-1}}$, we obtain $M_{\rm BH}=3.6^{+5.8}_{-1.7}\times10^7 M_\odot$.
We note that the broad H$\beta$ luminosity relative to the optical continuum is higher than expected from the relation observed in the local universe \citep{greene2005bhmass}, a trend generally seen in the broad-line emitters from our sample (X.~Cai et al.\ in preparation).
The same trend has also been reported for other high-redshift AGN samples \citep[e.g.,][]{degraaff2025lrd, yanagisawa2026emissionlineluminosity,  onoue2026shellqs}.
We therefore adopt the continuum-based virial mass calibration rather than one based on the broad Balmer luminosity.
Applying the quasar bolometric correction $L_{\rm bol}=9.26~L_{5100}$ \citep{richards2006bolometric}, we estimate an AGN bolometric luminosity of $L_{\rm bol}=5.0_{-1.1}^{+1.4}\times10^{44}~{\rm erg~s^{-1}}$.
The corresponding Eddington ratio, $\lambda_{\rm Edd}\equiv L_{\rm bol}/L_{\rm Edd}$, is $\lambda_{\rm Edd}=0.11^{+0.10}_{-0.07}$.

\begin{figure}
    \centering
    \includegraphics[width=\linewidth, trim={30 15 10 10}, clip]{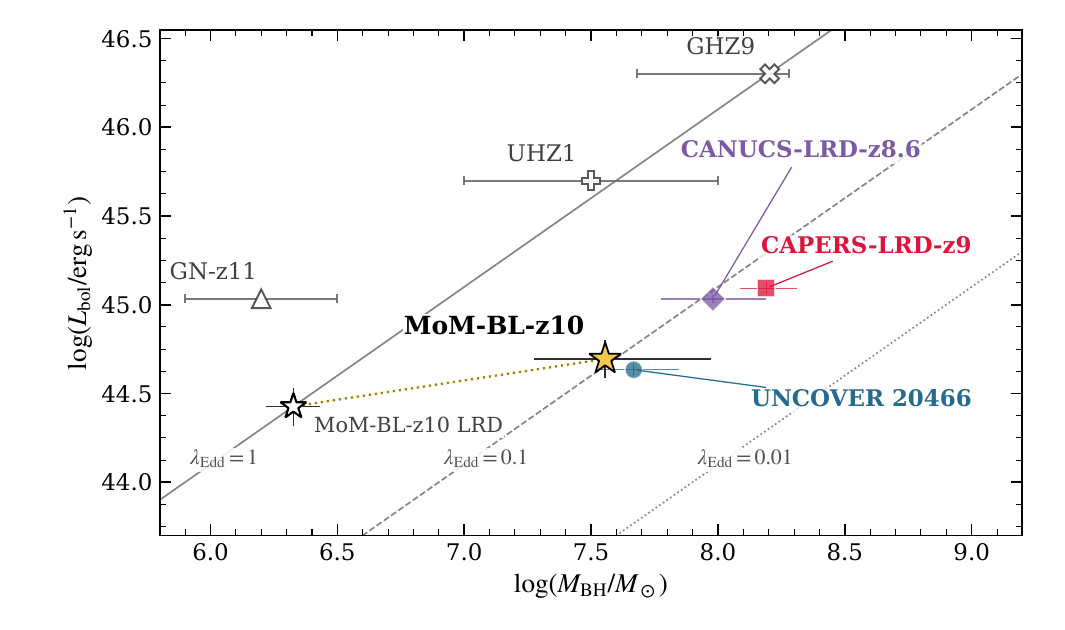}
    \caption{Black hole mass--bolometric luminosity distribution at $z\gtrsim8$.
    The yellow star shows \momblz{} with the black hole mass estimated using the H$\beta$-based calibration of \citet{kaspi2000bhmass} and the standard bolometric correction for  quasars  \citep[$L_{\rm bol}=9.26 L_{5100}$;][]{richards2006bolometric}.
    The white star shows the alternative LRD case for the same object, adopting the bolometric correction  of $L_{\rm bol, LRD}=5 L_{5100}$ \citep{greene2026lrdcalibration} and assuming an Eddington-limited accretion ($\lambda_{\rm Edd}=1$).
    In the LRD-based $M_\mathrm{BH}$ estimate, only the uncertainty of continuum luminosity is considered in the error budget, whereas the fiducial virial $M_\mathrm{BH}$ uncertainty includes those in both $L_{5100}$ and broad H$\beta$ FWHM.
    The black hole mass and bolometric luminosity of other $z\gtrsim8$ BL AGNs (CAPERS-LRD-z9 \citep{taylor2025capers}, UNCOVER 20466 \citep{kokorev2023lrd}, and CANUCS-LRD-z8.6 \citep{tripodi2025canucs}) are re-measured with the same procedure as for \momblz{} using the GH05 relation \citep{greene2005bhmass}.
    Open symbols show $z>9$ AGNs without broad line detection; GN-z11 \citep{maiolino2024gnz11}, UHZ1 \citep{bogdan2023bhmassevolution}, and GHZ9 \citep{napolitano2025ghz9}.
    We adopt $M_\mathrm{BH}$ and $L_\mathrm{bol}$ reported in the original papers.
    Diagonal lines indicate Eddington ratios of $\lambda_{\rm Edd}=1$, 0.1, and 0.01 from upper left to bottom right.}
    \label{fig:bhmass_luminosity}
\end{figure}

Given the red optical continuum and compact morphology of \momblz{} (Sec.~\ref{sec:momblz_phot}), we also consider an alternative interpretation motivated by recent LRD studies.
Adopting the LRD bolometric correction of $L_{\rm bol, LRD}=5L_{5100}$ \citep{greene2026lrdcalibration}, which is approximately half the standard value for quasars, we obtain $L_{\rm bol, LRD}=2.7^{+0.7}_{-0.6}\times10^{44}~{\rm erg~s^{-1}}$.
Recent LRD models suggest that SMBHs may undergo near- or super-Eddington accretion during the LRD phase  \citep{naidu2026lrdengine,sunwq2026lrd,curtis2026lrd}.
Motivated by these studies, we consider an illustrative case of Eddington-limited accretion ($\lambda_\mathrm{Edd}=1$) for \momblz.
Combining this assumption with the LRD-specific bolometric luminosity,
we obtain $M_{\rm BH,LRD}=2.1^{+0.6}_{-0.5}\times10^6~M_\odot$, which is approximately 17 times smaller than our fiducial virial BH mass estimate.
The quoted mass uncertainty reflects only the uncertainty in $L_{5100}$.
Figure~\ref{fig:bhmass_luminosity} compares these two cases with other $z\gtrsim8$ AGNs in the SMBH mass – bolometric luminosity plane.

\subsection{Implications for Early Black Hole Growth} \label{sec:BHmass_growth}

\begin{figure*}[t!]
    \centering
    \includegraphics[width=0.7\linewidth, trim={10 20 0 20}, clip]{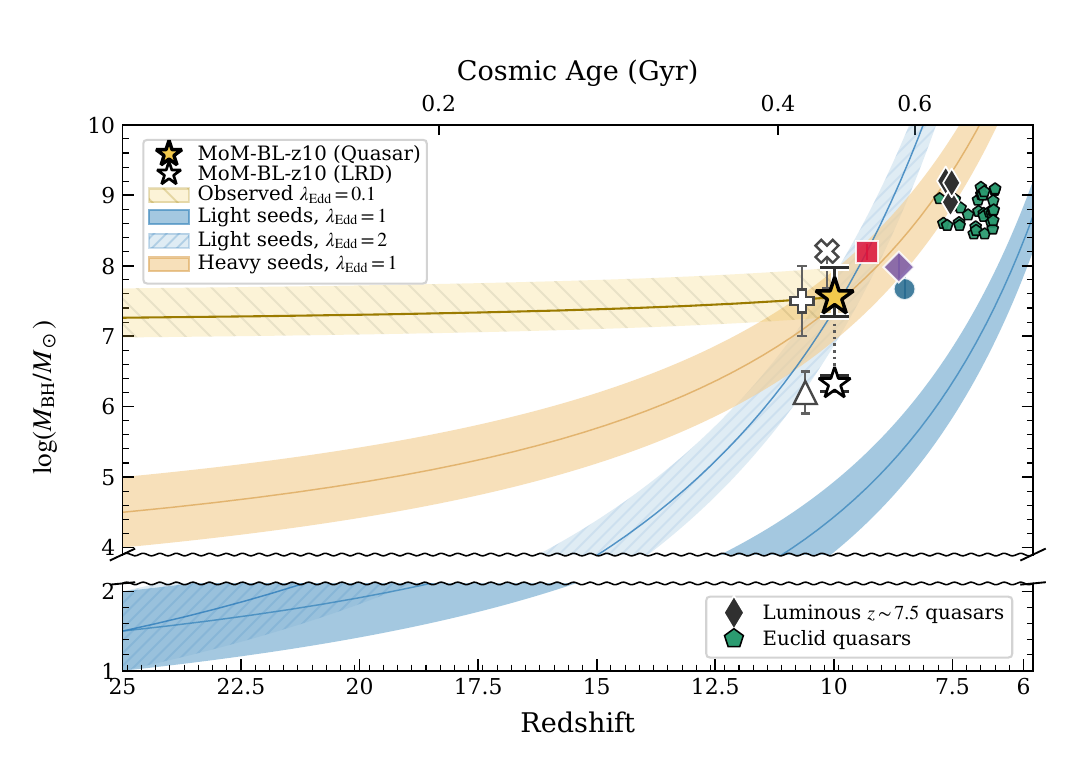}
    \caption{Illustrative black hole growth tracks of \momblz{} as a function of redshift assuming continuous accretion at fixed Eddington ratios with 10\%\ radiative efficiency.
    The yellow star shows the fiducial virial SMBH mass of \momblz{}, while the white star shows the LRD-based SMBH mass estimate assuming Eddington-limited accretion.
    The yellow stripe shows the backward extrapolation of the fiducial BH mass, assuming the observed Eddington ratio of $\lambda_\mathrm{Edd}=0.1$.
    The orange and blue stripes show the expected growth curves from a heavy ($10^{4-5}~M_\odot$) and a light (10--100~$M_\odot$) seed BH, respectively, assuming continuous Eddington-limited accretion ($\lambda_{\rm Edd}=1$) beginning at $z=25$.
    The widths of these stripes correspond to their seed mass ranges.
    The lighter, hatched blue stripe shows the same light-seed track for a super-Eddington accretion case ($\lambda_{\rm Edd}=2$).
    Other symbols show known high-redshift quasars and AGNs in the literature; $z\sim 7.5$ luminous quasars \citep[diamonds;][]{wang2021j0313,banados2018j1342,yang2020j1007}, Euclid quasars \citep[pentagons;][]{yang2026}, and JWST AGNs shown in Figure~\ref{fig:bhmass_luminosity} with the same symbols.
    UHZ1 is shifted by $\Delta z=+0.6$ from its spectroscopic redshift for clarity.
    For the Euclid quasars, black hole masses are estimated from their bolometric luminosities assuming $\lambda_{\rm Edd}=0.25$, the median value of the comparably luminous $z\sim6$ quasars \citep{onoue2026shellqs}.
    }
    \label{fig:bh_growth_tracks}
\end{figure*}

The inferred SMBH mass of \momblz{} at $z=9.98$ provides a valuable constraint on early SMBH growth.
Figure~\ref{fig:bh_growth_tracks} illustrates the mass growth history of \momblz{} as a function of redshift, where we assume a seed formation epoch of $z=25$ and continuous accretion with a 10\%\ radiation efficiency \citep{yu2002epsilon,marconi2004epsilon}.
For our fiducial virial mass estimate of $M_\mathrm{BH}=3.6\times10^7~M_\odot$, growth from stellar-mass BH seeds ($M_{\rm seed}\sim10$--$100~M_\odot$; \citealt{madau2001, haiman2001, heger2003}) requires a time-averaged Eddington ratio of $\lambda_\mathrm{Edd}\approx2$.
This echoes the well-known challenge of explaining the existence of massive SMBHs at $z\gtrsim7$ \citep{mortlock2011,banados2018j1342,wang2021j0313,yang2026}, as is also indicated by the darker blue curve in Figure~\ref{fig:bh_growth_tracks}, where we assume $\lambda_\mathrm{Edd}=1$.
Indeed, under a heavy-seed scenario ($M_{\rm seed}\sim10^{4-5}M_\odot$), continuous Eddington-limited accretion can produce both the inferred SMBH mass of \momblz{}, and those of luminous quasars at $z\gtrsim7$ and  other known JWST AGN at $z\gtrsim8$ \citep{taylor2025capers, kokorev2023lrd, tripodi2025canucs}.
Such massive BH seeds may form through the direct collapse of pristine gas clouds \citep{lodato2006massivebh,latif2022massivebh} or through runaway stellar collisions in dense star clusters \citep{gieles2018sms,williams2026collisions,rantala2026sms}.

We note that the fiducial Eddington ratio of $\lambda_{\rm Edd}=0.11$ reflects the accretion state at the observed epoch and would be insufficient to explain the inferred BH mass if maintained since $z=25$ under our growth assumptions, as illustrated by the yellow  stripe in Figure~\ref{fig:bh_growth_tracks}.
It is therefore more plausible that \momblz{} experienced more rapid accretion in the past, potentially including episodes of near- or super-Eddington accretion \citep{li2024}.

Under the alternative LRD interpretation discussed in Section~\ref{sec:BHmass}, the inferred BH mass of \momblz{} decreases to $M_{\rm BH,LRD}=2.1\times10^6~M_\odot$, substantially relaxing the requirements for early BH growth.
Nevertheless, given the limited time available for BH growth, even this lower mass is still challenging to explain through continuous Eddington-limited accretion from stellar-mass BH seeds (Figure~\ref{fig:bh_growth_tracks}).

\section{Conclusions}
\label{sec:conclusions}
We report the discovery of \momblz{}, the highest-redshift broad-line AGN currently known at $z=9.98$, identified through our systematic search for broad-line AGNs at $6<z<10$ using archival JWST/NIRSpec PRISM spectra.
A full description of the survey will be presented in X.~Cai et al.\ (in preparation).
Our main conclusions are as follows:

\begin{enumerate}

    \item \momblz{} at $z=9.98$ is the highest-redshift object in our broad-line AGN sample, selected solely on the basis of NIRSpec/PRISM spectroscopy of 1,247 objects at $6.0 \leq z\leq 10.3$ without photometric pre-selection. The H$\beta$ emission-line profile of \momblz{} is well reproduced by a model consisting of narrow and broad Gaussian components, with the narrow component constrained by the [O~{\sc iii}] $\lambda4960$ line profile. The BIC favors this model over a narrow-line-only model with $\Delta {\rm  BIC}=-15.55$. The broad H$\beta$ component is detected at S/N $=3.87$, with $\mathrm{ FWHM}=3480^{+2040}_{-950}$ km s$^{-1}$.

    \item The AGN interpretation of \momblz{} is further supported by its compact morphology in F444W, and the detection of strong [O~{\sc iii}] $\lambda4960$ and [Ne~{\sc iii}] $\lambda3869$ emission.
    The high-excitation emission-line ratios place \momblz{} in the AGN-like regions of both the [O~{\sc ii}]-FLR and OHNO diagnostic diagrams, well separated from the loci of star-forming galaxies.

    \item The rest-frame UV and optical slopes of \momblz{} are consistent with a $z=10$ LRD, albeit with a large uncertainty in $\beta_{\rm opt}$ ($=0.94\pm0.83$).
    We also find a tentative detection (S/N $=3.3$) in MIRI F770W, which, if real, lends further support to its LRD-like nature.

    \item We estimate the AGN bolometric luminosity of \momblz{} as $L_{\rm bol}=4.95^{+1.39}_{-1.09}\times10^{44}~{\rm erg~s^{-1}}$, adopting the standard quasar bolometric correction.
    Using the rest-frame 5100~\AA\ continuum luminosity and broad H$\beta$ line width, we derive a virial BH mass of $M_{\rm BH}=3.59^{+5.79}_{-1.71}\times10^7~M_\odot$ and an Eddington ratio of $\lambda_{\rm Edd}=0.11^{+0.10}_{-0.07}$.
    Alternatively, adopting the LRD-specific bolometric correction and assuming Eddington-limited accretion ($\lambda_{\rm Edd}=1$) yields an illustrative BH mass estimate of $M_{\rm BH,LRD}=2.1\times10^6~M_\odot$.

\end{enumerate}

The broad H$\beta$ emission in \momblz{} lies near the long-wavelength limit of the extended NIRSpec/PRISM coverage, while broad H$\alpha$ would be challenging to detect with MIRI due to its lower sensitivity.
Thus, the redshift of \momblz{} may be close to the practical redshift limit for identifying faint AGNs through broad Balmer emission with JWST.
Extending systematic spectroscopic searches to larger samples with well-characterized selection functions will help establish how common these early AGNs are and provide further constraints on BH formation and growth.

Observed at $z=9.98$, when the Universe was only $470$ Myr old, \momblz{} represents the highest-redshift broad-line AGN with an H$\beta$-based virial BH mass estimate, bringing into sharper focus the long-standing challenge of understanding rapid SMBH growth in the early Universe.

\begin{acknowledgments}
This work is based on observations made with the NASA/ESA/CSA James Webb Space Telescope. The MIRI data were obtained from the Mikulski Archive for Space Telescopes at the Space Telescope Science Institute, which is operated by the Association of Universities for Research in Astronomy, Inc., under NASA contract NAS 5-03127 for JWST. The specific observations analyzed can be accessed via \dataset[https://doi.org/10.17909/sngr-7t73]{https://doi.org/10.17909/sngr-7t73}.
The NIRSpec data products presented herein were retrieved from the Dawn JWST Archive (DJA). DJA is an initiative of the Cosmic Dawn Center (DAWN), which is funded by the Danish National Research Foundation under grant DNRF140.

RPN acknowledges support from JWST-GO-5224.
JM acknowledges funding by the European Union (ERC, AGENTS, 101076224).
KM acknowledges support from the Waseda University Grant for Special Research Projects (Project Number 2025C-484).
AdG acknowledges funding by the European Union (ERC, FIRST-GIANTS, 101221926). AdG is supported by the Lise Meitner Excellence program of the Max Planck Society.
PD warmly acknowledges support from an NSERC discovery grant (RGPIN-2025-06182).
This work is supported by JSPS KAKENHI Grant Numbers JP23K13149, 26K00736, 26K07155, 26H02069, JP26K17200 and 26K07139.
This work has received funding from the Swiss State Secretariat for Education, Research and Innovation (SERI) under contract number MB22.00072, as well as from the Swiss National Science Foundation (SNSF) through project grants 200020\_207349 and 2000-1-243073.
XC thanks John D. Silverman and Takumi S. Tanaka for insightful discussions.

\end{acknowledgments}

\begin{contribution}
X. C. conducted the investigation and prepared the initial draft of the manuscript.
A. K.\ I., M. O., and Y. F. developed the initial idea and research goals.
A. K.\ I. provided primary supervision and mentorship.
M. O. provided additional supervision and methodological guidance, and took a leading role in reviewing and editing the manuscript.
S. A. contributed to the investigation and discussion of the results.
All authors contributed to reviewing and editing the manuscript.
\end{contribution}

\facilities{JWST(NIRSpec, NIRCam, MIRI)}
\software{Astropy \citep{2013A&A...558A..33A,2018AJ....156..123A,2022ApJ...935..167A},
\texttt{emcee} \citep{foremanmackey2013emcee},
\texttt{lmfit} \citep{newville2014lmfit},
Matplotlib \citep{hunter2007matplotlib},
NumPy \citep{harris2020numpy},
SciPy \citep{virtanen2020scipy}}

\bibliography{sample701}{}
\bibliographystyle{aasjournalv7}

\end{document}

%% file: mom_blz10_properties.tex
\begin{deluxetable*}{ll}[p]
\tabletypesize{\scriptsize}
\tablewidth{0pt}
\tablecaption{Observed and inferred properties of \momblz\label{tab:mom_bl_z10_properties}}
\tablehead{\colhead{Property [unit]} & \colhead{Measurement}}
\startdata
R.A. (ICRS) [dd:mm:ss]\tablenotemark{a} & 02:17:07.229\\
Decl. (ICRS) [dd:mm:ss]\tablenotemark{a} &-05:09:38.63 \\
Redshift & $9.9837\pm0.0009$ \\
$f_\nu({\rm F090W})$ [nJy] & $<20.8$\\
$f_\nu({\rm F115W})$ [nJy] & $<20.3$\\
$f_\nu({\rm F150W})$ [nJy] & $<16.3$\\
$f_\nu({\rm F200W})$ [nJy] & $31.5\pm4.6$  \\
$f_\nu({\rm F277W})$ [nJy] & $14.6\pm3.6$  \\
$f_\nu({\rm F356W})$ [nJy] & $9.9\pm3.4$  \\
$f_\nu({\rm F444W})$ [nJy] & $18.1\pm5.3$  \\
$f_\nu({\rm F770W})$ [$\mu$Jy] & $0.20\pm0.06$\\
$f_\nu({\rm F1280W})$ [$\mu$Jy] & $<9.79$ \\
$f_\nu({\rm F1500W})$ [$\mu$Jy]& $<5.20$ \\
$f_\nu({\rm F1800W})$ [$\mu$Jy] & $<6.18$ \\
$M_{\rm 1450}$ [AB mag] & $-19.37^{+0.14}_{-0.13}$  \\
$\beta_{\rm UV}$ & $-1.64\pm0.21$ \\
$\beta_{\rm opt}$ & $0.94\pm0.83$ \\
Broad H$\beta$ flux\tablenotemark{b} ~[$10^{-20}$ erg s$^{-1}$ cm$^{-2}$] & $93.4^{+24.9}_{-23.3}$ \\
Narrow H$\beta$ flux\tablenotemark{b} & $51.9^{+10.0}_{-10.3}$ \\
Broad H$\beta$ FWHM [km s$^{-1}$ ] & $3480^{+2040}_{-950}$ km s$^{-1}$ \\
Narrow-line FWHM [km s$^{-1}$ ] & $<527$ km s$^{-1}$ (95\%) \\
Broad H$\beta$ velocity offset \tablenotemark{c} & $604^{+449}_{-429}$  \\
Si\,{\sc ii}] $\lambda1817$ flux\tablenotemark{b} & $57.1\pm17.8$ \\
{[O\,{\sc ii}] $\lambda\lambda3727,3730$ flux\tablenotemark{b}} & $28.2\pm5.1$ \\
{[Ne\,{\sc iii}] $\lambda3870$ flux\tablenotemark{b}} & $27.1\pm5.4$ \\
{[Ne\,{\sc iii}] $\lambda3969$+H$\epsilon$ flux\tablenotemark{b}} & $19.0\pm5.2$ \\
H$\gamma$+[O\,{\sc iii}] $\lambda4364$ flux\tablenotemark{b} & $46.8\pm8.5$\\
{[O\,{\sc iii}] $\lambda4960$ flux\tablenotemark{b}} & $149.3\pm11.0$ \\
$L_{5100}$ [$10^{43}$ erg s$^{-1}$] & $5.4^{+1.5}_{-1.2}$ \\
$M_{\rm BH}$ [$10^7\ M_\odot$]& $3.6^{+5.8}_{-1.7}$ \\
$L_{\rm bol}$ [$10^{44}$ erg s$^{-1}$]\tablenotemark{d}  & $5.0_{-1.1}^{+1.4}$ \\
$\lambda_{\rm Edd}$\tablenotemark{d}& $0.11^{+0.10}_{-0.07}$ \\
$L_{\rm bol, LRD}$\tablenotemark{d} [$10^{44}$ erg s$^{-1}$] & $2.7^{+0.7}_{-0.6}$\\
$M_{\rm BH, LRD}$\tablenotemark{d} [$10^6\ M_\odot$] & $2.1^{+0.6}_{-0.5}$\\
\enddata
\tablecomments{Quoted uncertainties are $1\sigma$ errors or 16th--84th percentile intervals. 
The NIRCam photometry is measured within a $d=0\farcs5$ aperture and taken from the DJA catalog.
Photometric upper limits are given at $3\sigma$.
Positive broad H$\beta$ velocity offsets indicate redshifts relative to the systemic velocity defined by the narrow emission lines.}
\tablenotetext{a}{The coordinates are based on the NIRSpec source position. 
We confirm that the NIRCam position is consistent within $0\farcs026$ of this position. 
All MIRI fluxes and upper limits are measured using forced aperture photometry centered on these coordinates (see text).}
\tablenotetext{b}{Emission line flux in units of $10^{-20}$ erg s$^{-1}$ cm$^{-2}$. 
Those for Balmer and [\ion{O}{3}] emission lines are based on single Gaussian fits.
We also report those weaker lines with S/N $\geq3$, for which the line flux is derived by directly integrating the continuum-subtracted spectrum.}
\tablenotetext{c}{The broad-H$\beta$ velocity offset is given in km s$^{-1}$ relative to the narrow-line systemic redshift. Positive (negative) values indicate a redshift (blueshift).}
\tablenotetext{d}{We adopt $L_{\rm bol}=9.26L_{5100}$ for the quasar case \citep{richards2006bolometric}, while we use $L_{\rm bol}=5L_{5100}$ for the LRD case \citep{greene2026lrdcalibration}.
The SMBH mass for the LRD case ($M_\mathrm{BH, LRD}$ assumes $\lambda_{\rm Edd}=1$).}
\end{deluxetable*}